\documentclass[conference]{IEEEtran}
\IEEEoverridecommandlockouts
\usepackage{cite}
\usepackage{amsmath,amssymb,amsfonts}
\usepackage{algcompatible}
\usepackage{graphicx}
\usepackage{textcomp}
\usepackage{xcolor}
\usepackage{graphicx}
\usepackage[english]{babel}
\usepackage[utf8]{inputenc}
\usepackage{algorithm}
\usepackage{tikz}
\usepackage{arevmath}     
\usepackage[noend]{algpseudocode}
\def\BibTeX{{\rm B\kern-.05em{\sc i\kern-.025em b}\kern-.08em
    T\kern-.1667em\lower.7ex\hbox{E}\kern-.125emX}}
\newcommand\copyrighttext{%
  \footnotesize \textcopyright 2022 IEEE. Personal use of this material is permitted. Permission from IEEE must be
  obtained for all other uses, in any current or future media, including
  reprinting/republishing this material for advertising or promotional purposes, creating new
  collective works, for resale or redistribution to servers or lists, or reuse of any copyrighted
  component of this work in other works.}
\newcommand\copyrightnotice{%
\begin{tikzpicture}[remember picture,overlay]
\node[anchor=south,yshift=10pt] at (current page.south) {\fbox{\parbox{\dimexpr\textwidth-\fboxsep-\fboxrule\relax}{\copyrighttext}}};
\end{tikzpicture}%
}
\begin{document}

\title{P4Filter: A two level defensive mechanism against attacks in SDN using P4
}
\author
{\IEEEauthorblockN{Ananya Saxena}
\IEEEauthorblockA{\textit{Dept of CSE} \\
\IEEEauthorblockA{\textit{IIIT Naya Raipur}}\\
}
\and
\IEEEauthorblockN{Ritvik Muttreja}
\IEEEauthorblockA{\textit{Dept of CSE} \\
\IEEEauthorblockA{\textit{IIIT Naya Raipur}}\\
}
\and
\IEEEauthorblockN{Shivam Upadhyay}
\IEEEauthorblockA{\textit{Dept of CSE} \\
\IEEEauthorblockA{\textit{IIIT Naya Raipur}}\\
}
\and
 \IEEEauthorblockN{K. Shiv Kumar}
\IEEEauthorblockA{\textit{Dept of CSE} \\
\IEEEauthorblockA{\textit{IIIT Naya Raipur}}\\
}
\and
\IEEEauthorblockN{Venkanna U.}
\IEEEauthorblockA{\textit{Dept of CSE} \\
\IEEEauthorblockA{\textit{IIIT Naya Raipur}}\\
}
}

\maketitle
\copyrightnotice
\begin{abstract}

The advancements in networking technologies have led to a new paradigm of controlling networks, with data plane programmability as a basis. This facility opens up many advantages, such as flexibility in packet processing and better network management, which leads to better security in the network. However, the current literature lacks network security solutions concerning authentication and preventing unauthorized access. In this work, our goal is to avoid attacks in a two level defense mechanism (P4Filter). The first level is a dynamic firewall logic, which blocks packets generated from an unauthorized source. The second level is an authentication mechanism based on dynamic port knocking. The two security levels were tested in a virtual environment with P4 based switches. The packets arriving at the switch from unknown hosts are sent to the controller. The controller maintains an ACL using which it assigns rules for both the levels to allow or drop the packets. For port knocking a new random sequence is generated for every new host. Hosts can only connect using the correct sequence assigned to them.The tests conducted show this approach performs better than the previous P4 based firewall approaches due to two security levels. Moreover, it is successful in mitigating specific security attacks by blocking unauthorized access to the network.
\end{abstract}


\section{Introduction}
The introduction of software defined networking (SDN)\cite{open_networking_foundation_2012} has led to networking paradigms becoming more flexible. The flexibility allows for more significant innovation and minimizes dependency on equipment manufacturers for inducing network changes. The increased programmability permitted by SDN enabled any client to develop and try their custom network algorithms without having to rely on manufacturers. It was still limited in its capabilities as OpenFlow\cite{hu2014survey} was a bottleneck in the programmability of network devices. To address such limitations, a Programming Protocol independent Packet Processor (P4) \cite{bosshart2014p4} was introduced which utilizes the concept of data plane programmability. This means rather than a dumb device at the data plane, and we use an intelligent switch known as P4 switch. The P4 program running on the switch helps to take various decisions without transferring packets to the controller. As a result, the switch can take action and enforce some control on packets. Besides giving control to the data plane, it also increases the execution speed as controller involvement is somewhat reduced. Due to the control statements and match action tables used in P4, we can block certain packets that violate the network policies to increase network security.

Networks, including SDN, suffer from various threats. Although moving to software defined networking gives a central control and greater flexibility, many security related concerns remain. The switches and hosts are thus vulnerable to attacks. In addition to this, attacks such as  Man in the Middle Attack, IP spoofing attack, DDoS attacks are possibly hampering the CIA (Confidentiality, Integrity, Availability) principle of network security. These problems occur due to the lack of well designed firewall and authentication mechanisms in SDN. Most of the current state of the art mechanisms in this regard work on traditional networking paradigms, thus, they offer very little flexibility. Also, there is no standard mechanism available that offers complete authentication and security in SDN.

The primary motivation behind the proposed solution is that the networking systems suffer from various security issues such as the Man in the Middle attack, IP spoofing, etc. Man in the middle attack\cite{al2020improving} can breach the confidentiality and integrity of data. It is due to a lack of proper authentication mechanisms and firewall systems. IP spoofing attack\cite{zhang2017towards} is one in which the attacker tries to mimic the IP address of some authorized host and attempts to breach the firewall. In the absence of a proper authentication mechanism, the impersonation and IP spoofing attack becomes easier. This can lead to information leaks in systems, which can, in turn, comprise confidential data to unauthorized persons. Furthermore, it can also lead to failures in data sensitive networks as any breach may cause faulty and manipulated data towards data centric applications which support the systems. Due to the shortcomings mentioned above, there is an increased possibility of failures in critical systems affecting many users. DDoS attack \cite{bawany2017ddos} \cite{hameed2018sdn} is also possible in SDN, leading to jamming of the nodes, hence rendering them inaccessible. The traditional port knocking approaches use the same port knocking sequence for all hosts to connect to a network. Therefore, the existence of a single port knocking sequence can lead to a compromised system\cite{von2015detecting} even with a single compromised host which is not ideal in networks.
    
The current paper focuses on providing a novel security mechanism for networking devices. We propose a two level security approach, P4Filter, in which the first level is the functionality of the P4 based dynamic firewall. It uses the concepts of both the stateful and stateless firewall. Level two of the mechanism is authentication using dynamic port knocking. \textbf{The significant contributions of this research are as follows:}
\begin{itemize}
\item A P4 based approach to block packets from hosts which are considered potential threats is proposed. It uses the match action tables to block specific hosts based on source and destination IP and MAC addresses. This acts as the first line of defence. An ACL list is also maintained by the controller using which it can block specific hosts who are potential threats, rather than allocating it a port knock sequence.
\item An authentication approach is also proposed so that hosts can establish their authenticity that is designed using dynamic port knocking.  In this, hosts trying to develop a connection should knock correct ports in sequence to connect, acting as the second line of defense.
\item Further, to keep the solution dynamic, the port knocking sequence is assigned by the controller based on P4Runtime whenever a new host tries to connect for the first time. The two-level approach supplemented by the controller ensures that the system can mitigate many security attacks that can otherwise prove to be harmful to the network.
\end{itemize}

The paper is structured as follows. Section II of the paper gives a review of the existing approach and research gap. Further, the methodology for P4Filter is described in Section III. Furthermore, Section IV gives the implementation details and detailed result analysis. Finally, the paper ends with Section V which provides a conclusion and mentions areas of future research.

\section{Related Works}
The data plane programmability and ease of doing experiments have led to many proposed ideas and papers. Several of these ideas address various types of concerns that arise in the field of network security. This section looks at such works.

\subsection{Previous state of the art approaches}

Some work has been done in the field of security in SDN in recent times. Here we look at a brief overview of the recent works.

Pakapol et al. \cite{krongbarameeStateful} successfully implemented a stateful firewall using Open vSwitch. They show how SDN stateful firewall work to reduce the overhead encountered in SDN switches. Datta et al. \cite{datta2018p4guard} modeled a stateful P4 based firewall known as P4Guard. The firewall works based on the policies predefined in the controller, with the policies pushed to the data plane table by the controller. Further, Cao et al. \cite{cao2019cofilter} proposed CoFilter, a stateful firewall that provides speed up and increased efficiency over traditional implementations by using a hash to compress the connection state.

Zaballa et al. \cite{zaballa2020p4knocking} have explored the use of port knocking on a P4 switch. The paper demonstrates the use of registers to track source IP addresses and the use of CRC hash of source to do the same. They also present an implementation that relies mainly on the controller for its functioning.


Almaini et al. \cite{almaini2021lightweight} explored the use of a ticketing mechanism in P4 switches using Port Knocking such that the traffic is forwarded only if the sender has a valid ticket. The predefined nodes have a ticket by default, while new nodes can obtain a ticket by completing a successful authentication via port knocking.

\subsection{Problems not addressed by previous solutions}
\begin{itemize}
    \item All the previous approaches employ a single solution and not a combination of all. Not using an authentication mechanism suffers from attacks like a Man in the Middle and other security threats. 
    \item No solution uses dynamic port knocking for authentication. The solution, which only uses a stateless Firewall, is susceptible to IP spoofing attacks. 
    \item The solution using simple port knocking with the same port knock sequence for all hosts may not keep the sequence confidential. However, this implies that the person knowing the correct sequence can intentionally or unintentionally leak the port knock sequence, thus hampering the overall security.
\end{itemize}

\section{Methodology}
\begin{figure}[h]
\includegraphics[width=\linewidth]{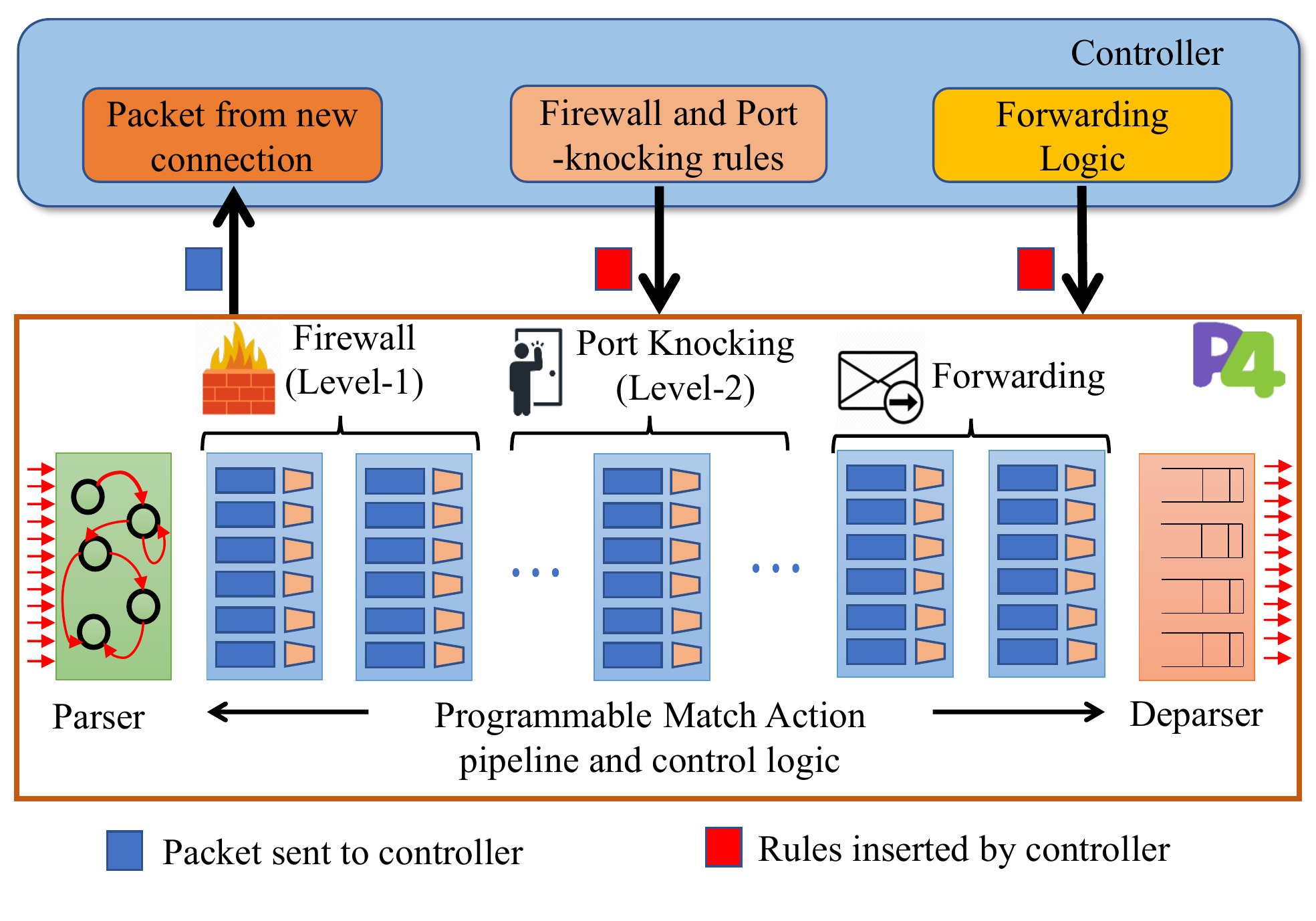}
\caption{P4Filter system architecture}
\label{fig:methodology}
\end{figure}

P4Filter is a two-level filtering approach to ensure only authorized hosts can communicate in the network. It also protects the network from unwanted traffic. The system is implemented in three modules as shown in Fig. \ref{fig:methodology}. The module-1 is the first Level filtering based on firewalls that can drop/allow the packets based on the sender and receiver addresses. Further, module-2 is second-level filtering. It is an authentication mechanism that blocks unauthorized hosts from connecting with the network using dynamic port knocking. In this port knocking, the order is different for different hosts. If the same device connects next time, it has to use the same port order allocated beforehand. If the packet is allowed by both levels, it is then forwarded using the forwarding logic specified in the Match-Action tables, which serves as the third module, as shown in Fig. \ref{fig:methodology}. 

\subsection{Packet processing}
There are a total of three main modules in packet processing. This processing happens in the P4 switch according to the compiled P4 program to the switch. It has two security modules and one module for forwarding. The security module is further divided into two levels for filtering the packets. Level-1 filtering focuses on a firewall approach, whereas level-2 filtering is on dynamic port-knocking. The details of sub-modules are described in the following subsections. After the packet is allowed by security modules, it is forwarded using the forwarding module.
\subsubsection{Level-1 Filtering - Firewall}
This module consists mainly of using firewalls as the first line of defense against attacks. It makes use of two types of firewalls stateless and stateful firewalls. First, it filters the packets by dropping packets that do not fulfill specific criteria. For example, those packets are dropped from the source with an IP address not known by the system, or those packets are dropped in which first communication is established by external hosts. The following are the two firewalls that were used and described in the following subsections.
\paragraph{Stateless Firewall}
The first part of Level-1 filtering is a stateless firewall. Stateless firewalls use information such as destination IP address, source IP address, and other parameters like MAC addresses to figure out whether a given packet poses any threat or not. In traditional networks, these firewalls are implemented using Access Control Lists (ACL). However, implementing a Stateless firewall is as simple as defining match action rules to drop the malicious/unknown packets in SDN. Algorithm \ref{alg:stateless} illustrates the implementation used in the SDN approach.

\begin{algorithm}
\caption{Stateless Firewall}
\label{alg:stateless}
\begin{algorithmic}
\Procedure{Stateless}{pkt\_in}
        \If{pkt\_in.SrcIP in check\_IP}
            \State check\_SrcMAC()
            \If{pkt\_in.SrcMAC in check\_MAC()}
            \State update\_check\_allow()
            \Else
            \State mark\_to\_drop()
            \EndIf
            \Else
            \State mark\_to\_drop()
        \EndIf
\EndProcedure

\end{algorithmic}
\end{algorithm}

In P4Filter, if a stateless firewall does not drop the packet and a match is found, then it is forwarded for further processing to a stateful firewall. Suppose no match is found, but no drop rule is found either, in that case, it is sent to the controller. The controller uses the ACL list to either drop the packet or first install forwarding rules and then send data to the host containing information about port knocking order. 

\paragraph{Stateful Firewall}
For the second part of Level-1 filtering, P4Filter uses a stateful firewall. By default, every external traffic is blocked while the internal hosts are still allowed to make external requests. In addition, an external host is allowed if initially any request was made from the internal host. Algorithm \ref{alg:stateful} explains the implementation of the stateful firewall used in P4Filter.
\begin{algorithm}
\caption{Stateful Firewall}
\label{alg:stateful}
\begin{algorithmic}
\Procedure{Stateful}{pkt\_in}
    \If{pkt\_in in internal\_port}
        \State check\_SYN\_Flag()
        \If{pkt\_in.syn == 1}
        \State update\_flow\_register()
        \State forward()
        \Else
        \State forward()
        \EndIf
        \Else
        \State check\_known\_flows
        \If{flag == 1}
        \State forward()
        \Else
        \State drop()
        \EndIf
    \EndIf
\EndProcedure
\end{algorithmic}
\end{algorithm}
The implementation of a Stateful firewall uses two Bloom filters \cite{byalibloom} \cite{geravand2013bloom} to maintain the list of flows that were sent from inside the network. Two filters are used to account for errors arising due to the probabilistic nature of the Bloom filter. For a packet to make through from outside, its server must have gotten a request from inside the network. When a packet is first sent from inside, the switch matches the input and output ports using the check\_ports table to see if it is coming from the internal network.

If the switch gets a match,  then 'direction' is set to Zero, and a hash is calculated using source IP address(i.e., from internal network), destination IP address, source port, and the destination port. Next, it is checked if the TCP syn bit is set or not, If it is set to zero, then the hash is written in the two bloom filters so that the next time a packet/reply comes from outside, it will allow the packet.

If the switch does not match(i.e., the packet is from outside), then the 'direction' bit is set to one, and a hash is calculated using the destination IP address, source IP address, destination port, and source port. Afterward, this hash is matched to see whether it is present in the two bloom filters or not. If it is found in both the bloom filters, then the packet can pass to the node in the internal network.

\subsubsection{Level-2 Filtering - Port Knocking}
Level-2 of filtering uses dynamic Port Knocking for authentication. In simple port knocking, a host needs to send TCP SYN packets known as port knocks to ports in correct predefined sequences. Using port knocking, the hosts with prior knowledge of the sequence can establish a connection that was otherwise not allowed to connect. To illustrate the idea, suppose a host knocks port sequence 2222, 3333, 4444(in that order), then if the order was correct, it is allocated port 22 to send the packets. The above mentioned scenario is depicted in Fig. 2. The problem with this approach is that it uses the same port knock sequence for every host that tries to connect. In our approach, we slightly modify port knocking. As soon as the new host tries to establish a connection, the controller is informed. The controller then allots a port knock sequence valid for a host with the particular IP address only. Although the sequence alloted is different for different hosts, the process of authentication remains same as shown in Fig. 2. Algorithm \ref{alg:portknock} explains how dynamic port knocking works in P4Filter.
\begin{figure}[h]
\includegraphics[width=\linewidth,height=7.6cm]{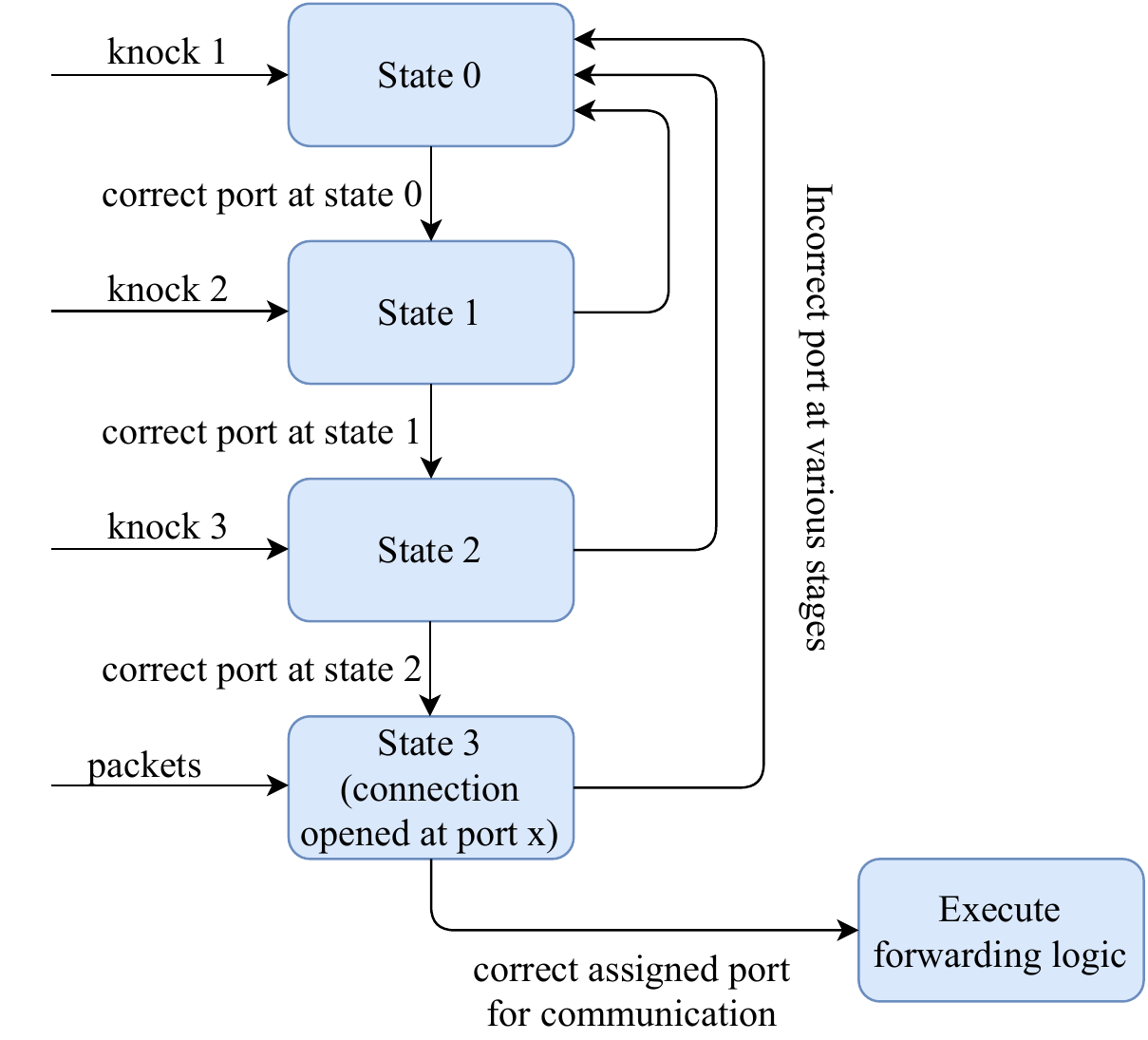}
\label{fig:portknock}
\caption{Illustration of state transition diagram for port knocking}
\end{figure}

\begin{algorithm}
\caption{Dynamic Port Knocking}
\label{alg:portknock}
\begin{algorithmic}
\Procedure{PortKnock}{packet in}
    \State Receive packet at an ingress port.
    \State Do table lookup on present\_table
        \If{the match is not found}
        \State Send the packet to the controller through CPU port.
        \State Check the host against ACL list on  controller
            \If{host not present in ACL list}
            \State Install rules to drop the packet in the switch. 
            \Else 
            \State Add host to present\_table.
            \EndIf
        \Else
        \State Check the port knocking sequence state
        \State Update the state
        \EndIf
        \If{port knock sequence is at the state three}
            \State Assign a destination port to the host
            \State Forward the packet using ipv4\_forward table
        \EndIf
\EndProcedure

\end{algorithmic}
\end{algorithm}
\subsubsection{Forwarding logic}
Once the packet has passed through all filtering levels, it is finally forwarded using rules in the Match-Action table. The Source and Destination addresses are located in the Match-Action table, and TTL value will be decremented. Further, the packet is placed on the appropriate output port of the switch obtained from the Match-Action table.

\subsection{Controller Communication}
Whenever a packet comes to switch, first the match action table \textit{present table} is checked to see if it is the first time that communication is happening with that particular host. If no entry for the host IP address is present in the present table, then the packet is sent to port 55 of the switch to reach the controller. The controller maintains an ACL list (as shown in Fig. 3) through which it knows whether to allow or block that host. 

\begin{figure}[h]
\begin{center}
\includegraphics[width=7cm,height=2.3cm]{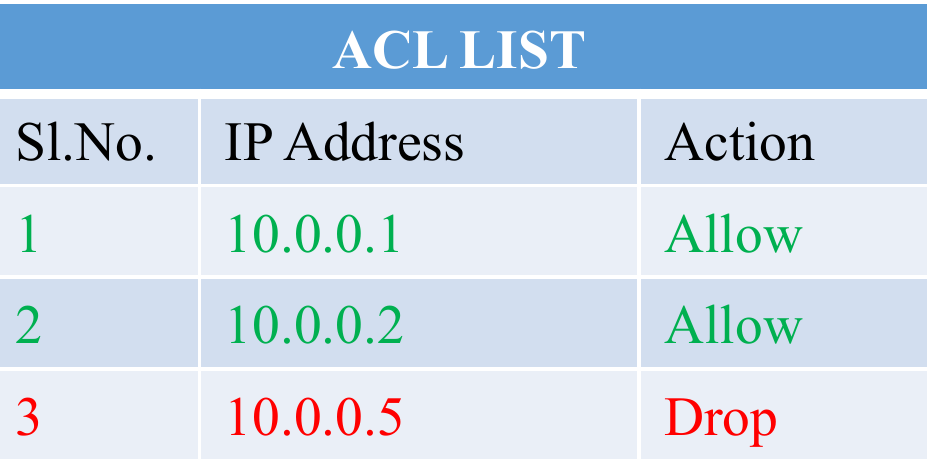}
\end{center}
\label{fig:acl}
\caption{A snapshot of P4Filter ACL list in the controller}
\end{figure}

If the entry is present and is says allow, a new port knocking sequence is allotted to the host. The necessary flow rules are inserted in the match action table of the switch so that the packet can be sent to appropriate hosts if the sender follows the port knocking pattern allotted to it. If the entry for the sender's IP is not present in the ACL list or entry is present which says drop, only the \textit{current table} rules are inserted so that the switch knows that subsequent packets from that host are not sent to the controller. The rules also set a parameter that communicates with the switch that the packets coming from that particular host should be dropped. By implementing this, we can have a system where authentication is done using port-knocking, and only hosts allowed by the network administrator are allowed to authenticate. Thus we can prevent IP spoofing attacks to a certain extent due to different port knocking sequence allotted to different hosts. Also, since every host has a unique port knocking sequence, the sequence leaked by one host cannot simply be used by the attacker. 


\section{Implementation and Result Analysis}
\subsection{Environment}
To demonstrate of our solution, a  network topology is designed using Mininet and BMV2 switches. P4 is used to program the switches, and P4Runtime\cite{p42017p4} was used for installing flow rules in the switches using the controller. The hosts connected directly to a switch from the internal network of that particular switch. The packets are sent to the controller through CPU ports assigned to the switches. For packet generation from the various hosts, the Scapy library is used. The controller also maintains a JSON file that stores all the port-knocking sequences set by the controller.
\subsection{Topology}
\begin{figure}[h]
\includegraphics[width=\linewidth,height=5.4cm]{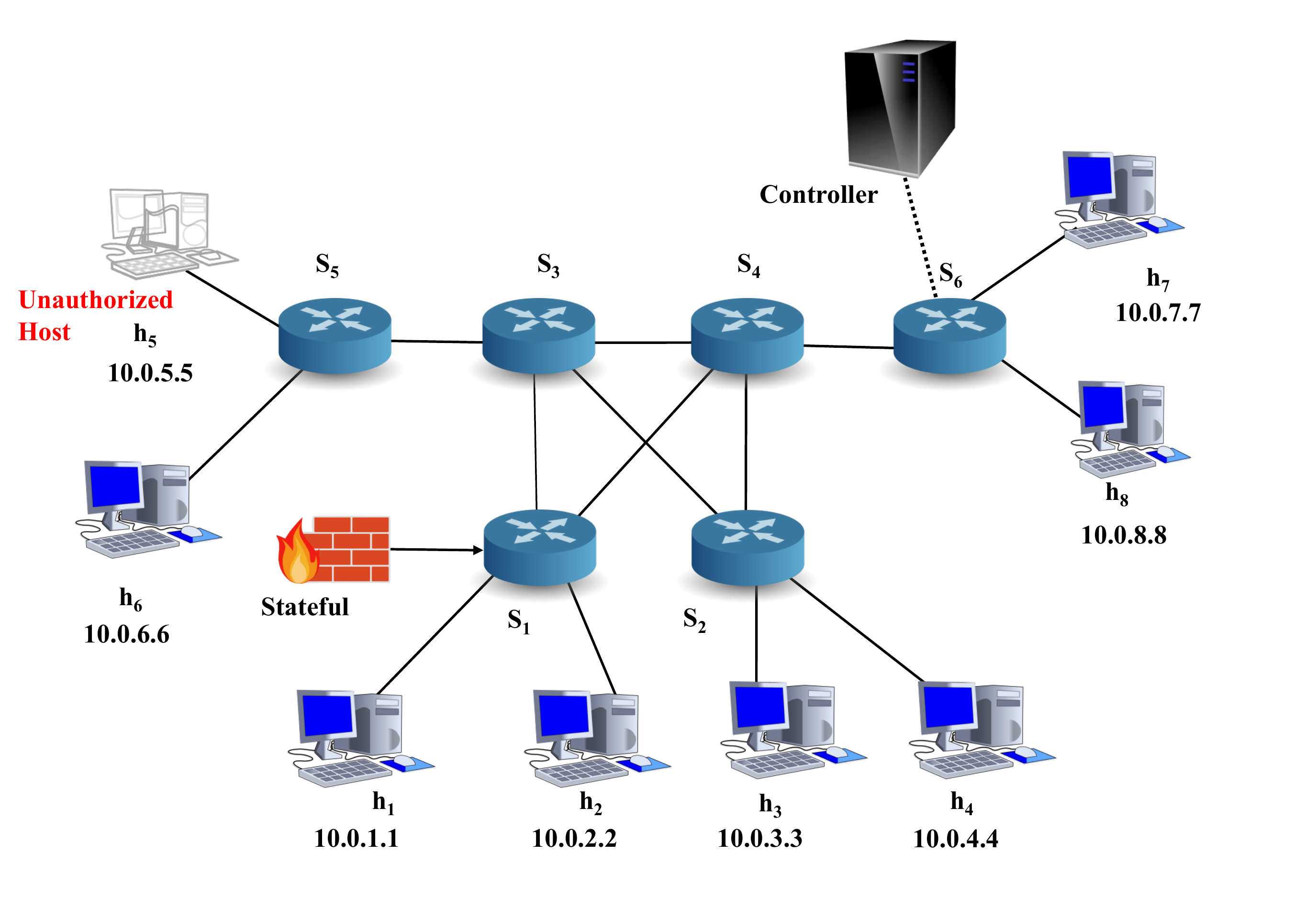}
\caption{P4Filter Implementation setup using Mininet}
\label{fig:topology}
\end{figure}

The topology was developed in Mininet and consisted of six switches and two hosts, each connected to four of the switches of the switch. Fig. \ref{fig:topology} shows the topology that is used for testing the approach. All the switches are BMV2 switches and run the P4 program on them, as discussed in the above section. Switch \textit{s1} runs a P4 program for stateful firewall, and Switch \textit{s2} runs a stateless firewall. In this topology, the switch \textit{s6} runs the port knocking with a stateless and stateful firewall. The rest of the switches run the primary program for just forwarding the packets based on flow. The simulation of packets from various hosts was carried using a packet generation library called Scapy.
\subsection{Stateful firewall}
\begin{figure}[h]
\includegraphics[width=\linewidth,height=1.8cm]{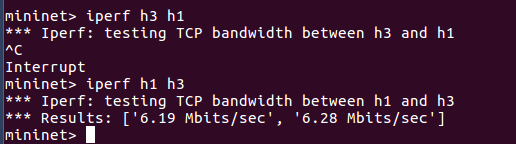}
\caption{A scenario for testing Stateful firewall using iperf command}
\label{fig:iperf}
\end{figure}

The P4 program for stateful firewall runs on switch \textit{s1}. It will only allow TCP connection to its internal network, i.e., host \textit{h1} and \textit{h2}, to the outside network only if the connection request is generated from the internal network. This is shown when the \textit{iperf} command is run from external host to internal host and from internal host to external host. We can see in the Fig. \ref{fig:iperf} that the \textit{iperf} is able to test bandwidth from \textit{h1} to \textit{h3}. It is also shown that no output is obtained when \textit{iperf} is run from \textit{h3} to \textit{h1}. The reason behind this is the fact that the switch blocks the communication through the stateful firewall. The stateful firewall works by checking whether the packet comes from an internal network or an external network. If the packet comes from an internal network, then the packet is forwarded according to the destination IP of the packet. In a bloom filter, the flow is added if the SYN flag is one in the TCP header, which later helps in finding if the request was initially generated from the internal network or not. If the packet comes from the external network, then the bloom filter is looked to find if the flow previously exists or not. If the flow exists, then the packet is forwarded; else, the packet is dropped. Fig. \ref{fig:iperf} shows the working of the stateful firewall using the \textit{iperf} command.
\subsection{Stateless firewall}
\begin{figure}[h]
\includegraphics[width=\linewidth,height=2.5cm]{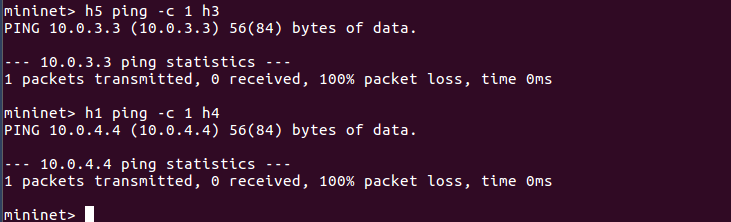}
\caption{Demonstration of Stateless firewall by pinging from unauthorized host}
\label{fig:statelessout}
\end{figure}

The P4 program for stateless firewall runs on switch \textit{s2}. It blocks host \textit{h5} based on IP address which is shown through the ping command in Fig. \ref{fig:statelessout}. Also, based on MAC Address, it blocks communication coming through switch \textit{s4} that is going in the internal network of the switch \textit{s2}. And thus, the switch can stop communication from any particular IP address or MAC address. Consequently, the network administrator has complete control of the firewall while it also reduces the need to have an extra device to act as a Firewall. This functionality can be used in two ways, i.e., the unauthorized hosts can be blocked by inserting drop rules for that particular host or switch and keeping the forwarding function as default for every other switch and host. The other approach by keeping the default action as a drop for all the devices and use forwarding actions only for the authorized devices can also be used. Unfortunately, the stateless firewall can be prone to IP spoofing attacks or impersonation attacks.
\subsection{Port Knocking}
In our implementation of we use port knocking to authorize the hosts trying to connect to the switch. As mentioned above the port knocking P4 program runs on the switch \textit{s6}. Whenever a packet is received from the host whose IP address is not present in the match action table of the switch \textit{s6}, then that packet is sent to the switch. The switch maintains an ACL list that contains the IP address of all the hosts which can be allowed to connect. A new port knock sequence is assigned for these hosts, and the required flow rules are inserted in the match action table. If the host's IP address is not present in the ACL list, then the rules to drop the packets from that particular IP address are installed. This is how it offers a complete authentication mechanism and implements the functionality of a stateless firewall. Afterward, whenever the host needs to connect, it will have to knock the correct sequence of ports assigned to it by the controller. If it does not use the correct sequence, then it will not be allowed to connect.
\begin{figure}[h]
\includegraphics[width=\linewidth,height=2.9cm]{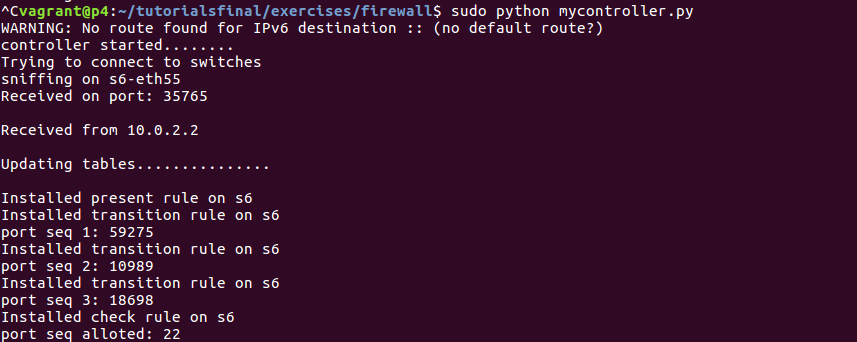}
\caption{A scenario representing allocation of port knocking sequence by controller}
\label{fig:controller}
\end{figure}
It is shown in Fig. \ref{fig:controller} that when the packet is received at the controller, then the controller first checks the ACL. If the source IP is present in the ACL list, then the port knocking sequence and all the corresponding rules to forward the packet to the appropriate port are also installed. For example, for the host with IP address "10.0.2.2," the allotted port knocking sequence is 59275, 10989, 18698, and the communication will happen after authentication using the port knocking sequence at port 22.

\begin{figure}[h]
\includegraphics[width=\linewidth]{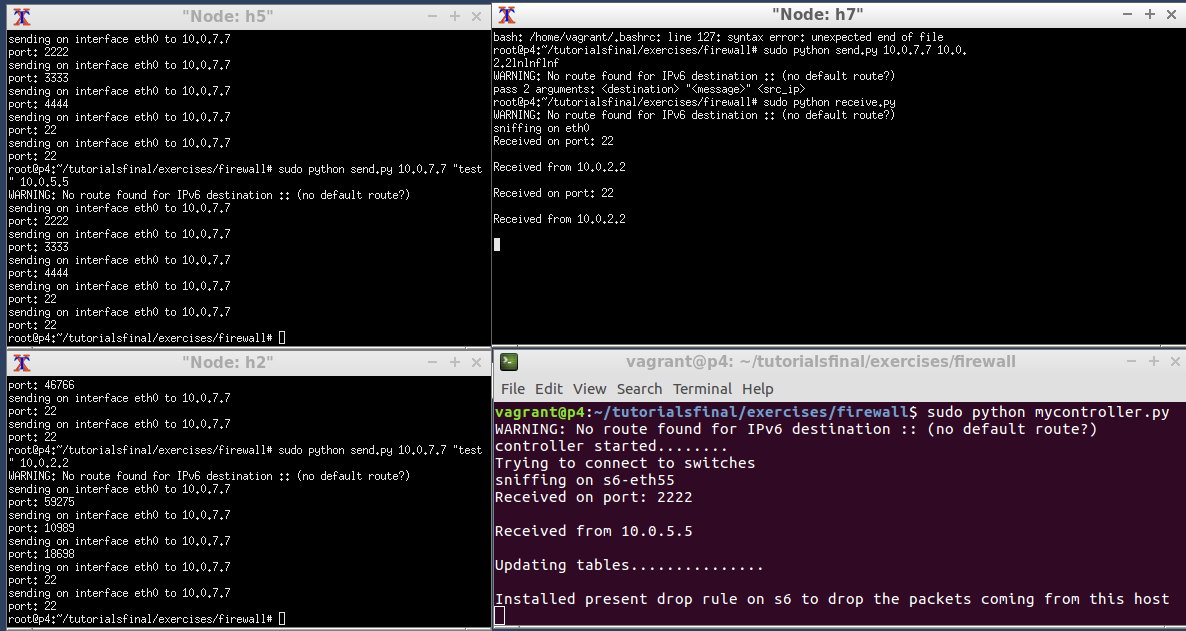}
\caption{Virtual simultaion of the security system}
\label{fig:host}
\end{figure}

Figure \ref{fig:host} shows that when the packet is received by the controller and the source IP address is not present in the ACL, the controller installs the rules to drop the packets. Also, the exact figure shows that when the rules are installed, and port knocking is done to open the connection, host \textit{h2} being authenticated properly can send the packets to host \textit{h7}. It is also shown the packets from host \textit{h5} are dropped due to the rules inserted by the controller.
\subsection{Comparison with the state of the art}
We compare the previous works explaining the limitations of the previously attempted solutions and how our implementation can give better results. The comparison is based on DDoS vulnerability, the flexibility of the approach, Protection from IP spoofing, and overall security achieved.



It can be observed that in the case of a DDoS Attack, port knocking \cite{zaballa2020p4knocking} performs best. It can be attributed to the authentication requirement, which does not allow packets from unauthorized hosts. P4Filter takes care of this issue by having a dynamic port knocking as the authentication mechanism. In terms of flexibility, P4Filter ranks high due to its dynamic nature and because of the flexible nature of P4. Our approach also prevents IP spoofing\cite{zhang2017towards} due to the authentication mechanism, whereas a simple stateless firewall can be bypassed through IP spoofing. As far as security goes, P4Filter performs fairly well as it prevents the unidentified nodes from reaching into the network and retains the connected node's identity through the port knocking sequence. Considering the above facts, we can easily conclude that the security provided by the individual approaches is not enough. P4Filter effectively resolves this problem.

\section{Conclusion and Future Work}
The two level security balances the downsides that arise due to either a stateless firewall or port knocking alone. It leads to a more reliable and robust security system. The proposed method, can effectively protect the network from various types of network attacks, including but not limited to IP spoofing attacks, Man in the middle attack, and DDoS attacks. The system is highly efficient, quite flexible, and offers the network administrator control to realign the functionalities according to the specific requirements. However, further research needs to be conducted to find the best way to transmit the allocated port knocking sequence to the host. Also, the system is currently tested in a virtual environment. The application of the system in real hardware can be explored, and the analysis of the results in real world applications could be quite fruitful.






\bibliographystyle{./bibliography/IEEEtran}
\bibliography{./bibliography/references.bib}

\begin{thebibliography}{10}
\providecommand{\url}[1]{#1}
\csname url@samestyle\endcsname
\providecommand{\newblock}{\relax}
\providecommand{\bibinfo}[2]{#2}
\providecommand{\BIBentrySTDinterwordspacing}{\spaceskip=0pt\relax}
\providecommand{\BIBentryALTinterwordstretchfactor}{4}
\providecommand{\BIBentryALTinterwordspacing}{\spaceskip=\fontdimen2\font plus
\BIBentryALTinterwordstretchfactor\fontdimen3\font minus
  \fontdimen4\font\relax}
\providecommand{\BIBforeignlanguage}[2]{{%
\expandafter\ifx\csname l@#1\endcsname\relax
\typeout{** WARNING: IEEEtran.bst: No hyphenation pattern has been}%
\typeout{** loaded for the language `#1'. Using the pattern for}%
\typeout{** the default language instead.}%
\else
\language=\csname l@#1\endcsname
\fi
#2}}
\providecommand{\BIBdecl}{\relax}
\BIBdecl

\bibitem{open_networking_foundation_2012}
G.~Tank, A.~Dixit, A.~Vellanki, and D.~Annapurna, ``Software-defined
  networking: The new norm for networks,'' \emph{3rd National Conference on
  Recent Innovations in Science and Engineering}, May 2017.

\bibitem{hu2014survey}
F.~Hu, Q.~Hao, and K.~Bao, ``A survey on software-defined network and openflow:
  From concept to implementation,'' \emph{IEEE Communications Surveys \&
  Tutorials}, vol.~16, no.~4, pp. 2181--2206, 2014.

\bibitem{bosshart2014p4}
P.~Bosshart, D.~Daly, G.~Gibb, M.~Izzard, N.~McKeown, J.~Rexford,
  C.~Schlesinger, D.~Talayco, A.~Vahdat, G.~Varghese \emph{et~al.}, ``P4:
  Programming protocol-independent packet processors,'' \emph{ACM SIGCOMM
  Computer Communication Review}, vol.~44, no.~3, pp. 87--95, 2014.

\bibitem{al2020improving}
A.~Al-Hayajneh, Z.~A. Bhuiyan, and I.~McAndrew, ``Improving internet of things
  (iot) security with software-defined networking (sdn),'' \emph{Computers},
  vol.~9, no.~1, p.~8, 2020.

\bibitem{zhang2017towards}
C.~Zhang, G.~Hu, G.~Chen, A.~K. Sangaiah, P.~Zhang, X.~Yan, and W.~Jiang,
  ``Towards a sdn-based integrated architecture for mitigating ip spoofing
  attack,'' \emph{IEEE Access}, vol.~6, pp. 22\,764--22\,777, 2017.

\bibitem{bawany2017ddos}
N.~Z. Bawany, J.~A. Shamsi, and K.~Salah, ``Ddos attack detection and
  mitigation using sdn: methods, practices, and solutions,'' \emph{Arabian
  Journal for Science and Engineering}, vol.~42, no.~2, pp. 425--441, 2017.

\bibitem{hameed2018sdn}
S.~Hameed and H.~Ahmed~Khan, ``Sdn based collaborative scheme for mitigation of
  ddos attacks,'' \emph{Future Internet}, vol.~10, no.~3, p.~23, 2018.

\bibitem{von2015detecting}
F.~von Eye, M.~Grabatin, and W.~Hommel, ``Detecting stealthy backdoors and port
  knocking sequences through flow analysis,'' \emph{PIK-Praxis der
  Informationsverarbeitung und Kommunikation}, vol.~38, no. 3-4, pp. 97--104,
  2015.

\bibitem{krongbarameeStateful}
P.~Krongbaramee and Y.~Somchit, ``Implementation of sdn stateful firewall on
  data plane using open vswitch,'' \emph{15th International Joint Conference on
  Computer Science and Software Engineering (JCSSE)}, pp. 1--5, 2018.

\bibitem{datta2018p4guard}
R.~Datta, S.~Choi, A.~Chowdhary, and Y.~Park, ``P4guard: Designing p4 based
  firewall,'' \emph{IEEE Military Communications Conference (MILCOM)}, pp.
  1--6, 2018.

\bibitem{cao2019cofilter}
J.~Cao, Y.~Liu, Y.~Zhou, C.~Sun, Y.~Wang, and J.~Bi, ``Cofilter: A
  high-performance switch-accelerated stateful packet filter for bare-metal
  servers,'' \emph{28th International Conference on Computer Communication and
  Networks (ICCCN)}, pp. 1--9, 2019.

\bibitem{zaballa2020p4knocking}
E.~O. Zaballa, D.~Franco, Z.~Zhou, and M.~S. Berger, ``P4knocking: Offloading
  host-based firewall functionalities to the network,'' \emph{23rd Conference
  on Innovation in Clouds, Internet and Networks and Workshops (ICIN)}, pp.
  7--12, 2020.

\bibitem{almaini2021lightweight}
A.~Almaini, A.~Al-Dubai, I.~Romdhani, M.~Schramm, and A.~Alsarhan,
  ``Lightweight edge authentication for software defined networks,''
  \emph{Computing}, vol. 103, no.~2, pp. 291--311, 2021.

\bibitem{byalibloom}
P.~Byali, M.~Z.~S. Bevinahalli, and V.~Chavan, ``Bloom filter.''

\bibitem{geravand2013bloom}
S.~Geravand and M.~Ahmadi, ``Bloom filter applications in network security: A
  state-of-the-art survey,'' \emph{Computer Networks}, vol.~57, no.~18, pp.
  4047--4064, 2013.

\bibitem{p42017p4}
P.~L. Consortium \emph{et~al.}, ``P4 runtime,'' \emph{Website, https://github.
  com/p4lang/PI}, 2017.

\end{thebibliography}

\end{document}